\documentclass[floatfix,showkeys,showpacs,twocolumn,amsmath,amssymb,epsf,aps]{revtex4}
\usepackage{dcolumn}
\usepackage{epsfig}

\usepackage{graphicx}
\usepackage{longtable}
\usepackage{dcolumn}
\usepackage{bm}

\def\etal{{\em et al. }}
\def\abinit{{\it ab initio }}

\def\ligands{{[(C$_4$H$_9$)$_4$P$^{+1}$}]$_3$}

\def\nimol{[$\mbox{As}@\mbox{Ni}_{12}@\mbox{As}_{20}]^{-3}$ }
\def\nimolo{$\mbox{As}@\mbox{Ni}_{12}@\mbox{As}_{20}$ }

\def\nicage{As@Ni$_{12}$ }
\def\ascage{As$_{20}$ }

\def\cmin {cm$^{-1}$ }

\begin{document}

\title{ Electronic structure and rebonding in the onion-like \nimolo cluster}

\author{Tunna Baruah$^{1,4}$, Rajendra R. Zope $^2$, Steven L. Richardson$^{3,4}$, and Mark R. Pederson$^4$}
\affiliation{$^1$Department of Physics, Georgetown University, Washington DC, 20057}

\affiliation{$^2$School of Computational Sciences, George Mason University, Fairfax,
VA 22030}

\affiliation{$^3$Department of Electrical Engineering and Materials Science Research Center, Howard University, School of Engineering, Washington, DC 20059}

\affiliation{$^4$Center for Computational Materials Science, Naval Research 
Laboratory, Washington DC 20375-5345}

\date{\today}



\begin{abstract}
    We  present the first \abinit study of the geometry, electronic structure, 
charged states, bonding  and vibrational modes of the recently
synthesized fullerene-like \nimolo cluster which has icosahedral point 
symmetry [Science, {\bf 300}, 778 (2003)].
We show that the molecule is vibrationally stable and will be electronically
most stable in its -3 oxidation state in the condensed phase and in $-2$ state in 
the gas phase.
We examine the bonding in this unusually structured molecule from charge
transfer between atoms, infrared and Raman spectra, and charge density
isosurfaces.

\end{abstract}

\maketitle

Since the discovery that carbon can form fullerene-like structures of high symmetry 
\cite{smalley,KLFH}, there has been considerable speculation, both experimentally
and theoretically, as to 
whether symmetric fullerene-like clusters can be created from other elements 
\cite{metcar,sicr}. Solids comprised of such inorganic 
sub-units can have important applications in nanotechnology and materials science. An 
important step towards this goal was realized with the recent synthesis of the
\ligands \nimol salt by Moses \etal \cite{MFE} where the \nimol anion (Cf. Fig. 1a) is comprised of an icosahedral (I$_h$) \nicage cluster with an arsenic atom core
(Cf. Fig. 1b) encapsulated by a dodecahedral fullerene-like \ascage cage 
(Cf. Fig. 1c) to give an onion-skin-like \nimol cluster with icosahedral point symmetry.
As discussed by Moses \etal \cite{MFE}, clusters with this structure form the inner core
of a class of intermetallic compounds which were first identified by Bergman \etal
\cite{BWP}. Such cores might actually be responsible for the formation of quasicrystals.
The encapsulating \ascage is shows  only 
structural resemblance to a carbon-based fullerene. Electronically, the atoms in 
\ascage are $sp^3$-hybridized with three bonding orbitals along the surface of the 
dodecahedron and a fourth orbital containing a lone pair of electrons pointing outward 
from the cage. In the
carbon-based fullerenes the carbon atoms are $sp^2$-hybridized forming a strong 
covalently-bonded network of $\sigma$-bonds surrounded by a sheath of  
$\pi$-electrons in bonding orbitals.
 
The \nimolo cluster is fascinating in that 
it represents another example of what M\"{u}ller has called the ``beauty of symmetry" 
where the interpenetrating nature of the \ascage dodecahedron and the \nicage 
icosahedron yield a polyhedra of sixty triangular faces \cite{MUE}. Further, it 
demonstrates the first experimental confirmation of a completely non-carbon
fullerene-like dodecahedron, As$_{20}$.
At this stage, very little is known about the molecule and
the only theoretical study to date has 
calculated the molecular orbital energies 
with the semi-empirical extended H\"{u}ckel method \cite{MFE}.
\begin{figure}[b]
\epsfysize=1.6in
\epsfig{file=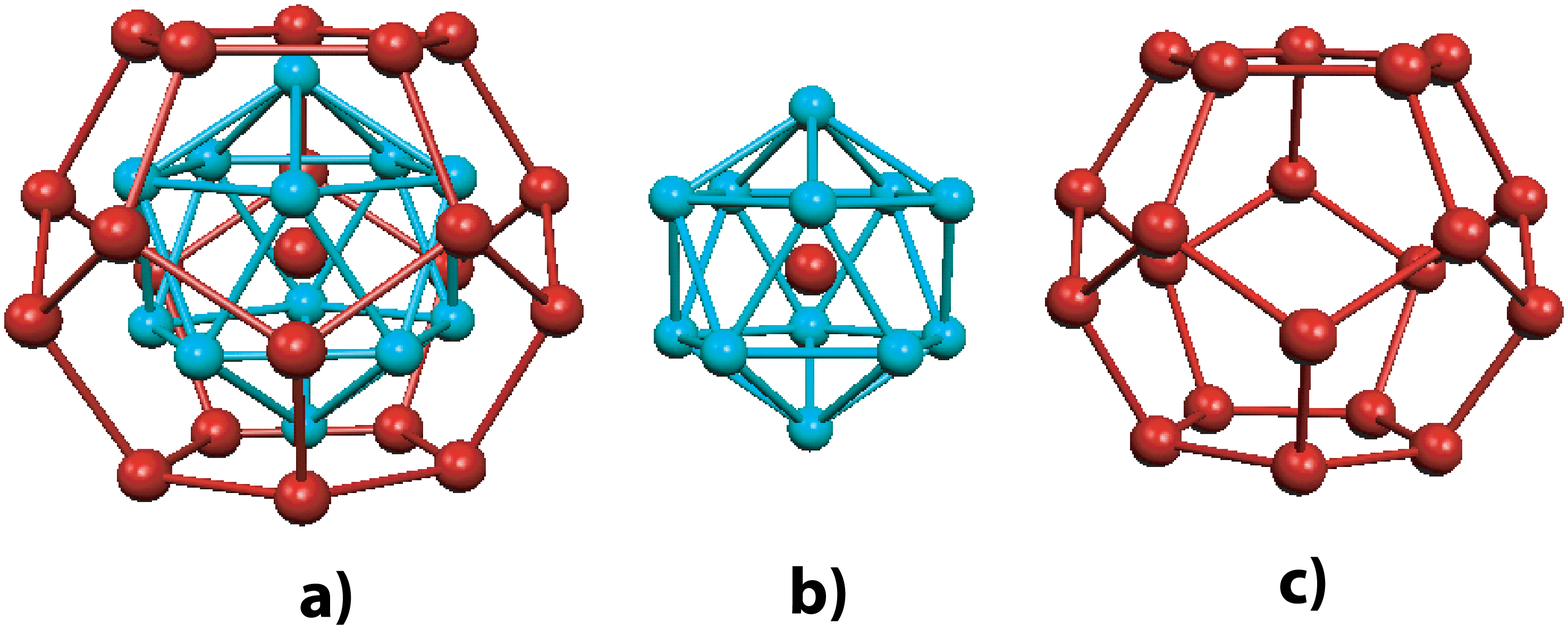,width=\linewidth,clip=true}
\caption{\label{fig1}  Structures of a)\nimolo, b) \nicage,
and c) \ascage .
}
\end{figure}
In this communication,  we present results of our first-principles density
functional theory (DFT)  studies on the isolated neutral and anionic states of 
the \nimolo  cluster and its sub-units, \ascage and \nicage, to determine their 
respective electronic structure, vibrational stability,
and chemical bonding patterns. In particular, our calculations
demonstrate that 
a rehybridization of the
the covalent As-As bonds
present in the isolated \ascage cage takes place to form new As-Ni bonds 
between the
inner As@Ni$_{12}$ cluster
and the outer \ascage cage.

Our DFT calculations  were performed
at the all-electron level with 
the generalized gradient approximation 
(GGA) \cite{PBE}  to describe the exchange-correlation effects. 
We have performed these calculations with T$_h$ symmetry but we find that
the optimized geometry has I$_h$ symmetry which was verified from the
positions of the atoms.  
All calculations have been performed using the 
NRLMOL package \cite{NRLMOL} which employs a Gaussian basis set where the exponentials are
optimized for each atom.

The optimized equilibrium geometries of the isolated \nimolo molecule along 
with its two  sub-units are shown in Fig. 1.  
For the free standing icosahedral \nicage\!, 
the core As-Ni distance is 2.36 \AA \, while the Ni-Ni
distance is 2.48 \AA. In the \ascage cage, the As-As 
distance is 2.5 \AA \, which is in good agreement with earlier calculated
results  \cite{As20}.
Upon formation of the \nimolo structure, the central As-Ni, Ni-Ni, and the outer cage
As-As distances increase
to 2.61,  2.74, and  2.8 \AA, \, respectively.
The Ni to outer layer As distance is 
2.43 \AA. The interatomic distances increase by about 10-12 \%  upon 
formation of the \nimolo molecule as compared to isolated sub-units.  
These bond lengths are in excellent
agreement with the experimental values 
(Cf. Table \ref{table1}).
We note that while our calculated bond lengths are for the neutral molecule,
the experimental values are for the molecule in the crystalline
environment where it  is in $-3$ charged state and the perfect I$_h$ symmetry
is weakly broken by standard crystal field effects and by the positively
charged ligands.
Our calculations have shown that 
the  structural changes induced by the excess charge are
small - from 0.4 to 1 \%. 
\begin{table*}
\caption{
The experimental and calculated values of bond lengths are presented  in \AA.
Note that (c) refers to the central As atom. The eigenvalues of the HOMO and the 
LUMO levels are given in Hartrees. The last column shows the Hubbard  U parameter
in Hartree/electron$^2$.}
 \label{table1}
\begin{ruledtabular}
\begin{tabular}{lccccccc}
 Method  & As - As  & Ni - Ni & As - Ni & As(c) - Ni &E$_{HOMO}$ &E$_{LUMO}$ & U     \\
\hline
 
 Expt.           & 2.75 &      & 2.40 & 2.56 &  \\
 \ascage         & 2.50 &      &      &      &-0.1974 & -0.1443  & 0.1052\\
 \nicage         &      & 2.48 &      & 2.36 &-0.1445 & -0.1384  & 0.1266\\
 \nimolo         & 2.80 & 2.74 & 2.43 & 2.61 &-0.1827 & -0.1789  & 0.1029\\
 \nimol          & 2.80 & 2.72 & 2.44 & 2.59 & \\
\end{tabular}
\end{ruledtabular}
\end{table*}

 The calculated atomization energy of the free-standing \ascage unit is 2.79 eV and that
of the \nicage is 3.05 eV. 
The \nimolo molecule has  atomization energy of  3.7 eV and 
is stable by about 26 eV with respect to its two sub-units. Its ionization potential
is 6.4 eV.

\begin{figure}
\epsfig{file=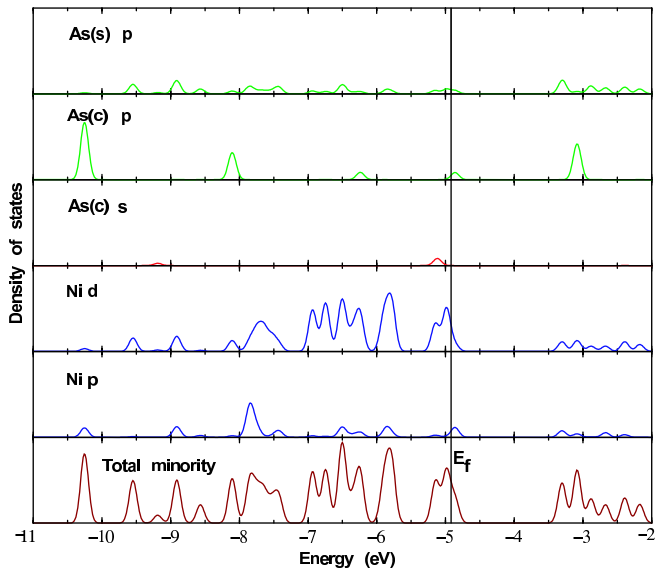,width=\linewidth,clip=true}
\caption{\label{fig2} The total and atom projected partial density
of states in arbitrary units of \nimolo of the minority channel. The three upper panel
shows the projected density on the central (c) and outer shell (s) As atoms. The Ni p
DOS is magnified for presentation purpose. }
\end{figure}
 The calculated total and atom projected density of states (DOS) of the
isolated neutral molecule are shown in 
Fig. \ref{fig2}. Since  the majority states are completely occupied,
we show the DOS for minority states responsible for chemical interaction.
Both the highest occupied molecular orbital (HOMO) and the lowest unoccupied
molecular orbital belong to the minority states. 
The HOMO has H$_u$ symmetry, while the LUMO has T$_{1u}$ symmetry.
The eigenvalues of HOMO and LUMO along with the Hubbard U values are
presented in Table \ref{table1}.
The neutral cluster has a spin magnetic moment of 3 $\mu_B$.
Reduction to the  $-3$  charged state 
quenches the magnetic moment and opens up a large HOMO-LUMO gap,
as is evident from the Fig. \ref{fig2}. 
This in turn stabilizes the molecule by 2.4 eV. 
The HOMO-LUMO gap in the  neutral and triply charged
state are
0.115  and  1.45 eV, respectively.
This observation is in accord with the magnetic susceptibility measurement
as well as with the H\"uckel model prediction \cite{MFE}. 
The DOS of the \nicage molecule (not shown) shows 
As s states in the excited levels which become occupied in the \nimolo complex.
The isolated \ascage is stable with a large HOMO-LUMO gap of 1.44 eV. 
In \nimolo, Ni d minority spin states are filled and 
the LUMO have a predominantly p character.  An analysis 
of the atom projected DOS  and the orbital densities shows 
that both the HOMO and LUMO are centered around the Ni  atoms.
Our calculation on $-1$ and $-2$  charged states show that the isolated molecule
is most stable in the $-2$ charged state. 
Although the electronic states are completely filled in the triply
charged anion, the electrostatic repulsion due to the extra
electrons makes the molecule energetically unfavorable in the gas phase.
The first and second electron affinities of the \nimolo are 3.47 and  0.74 eV
respectively. The accuracy of the electron affinities is subject to the
exchange-correlation functional used \cite{RKTS}. The low binding energy 
of the second electron makes it
susceptible to detachment in the laser desorption experiments.
This is consistent with the observation that the mass spectrum 
shows peaks for the singly charged anions \cite{MFE}.

\begin{table}[b]
\caption{
The charge within a sphere of radius 2.28 a.u. and 2.17 a.u. around
the As and
Ni atoms, respectively. The symbols (c) and (s) refer to the central
and outer shell.
}
\label{table3}
\begin{ruledtabular}
\begin{tabular}{lllll}
 Complex & Ni & As(c) & As(s)      \\
\hline
 \ascage &       &      & 31.06 \\
 \nicage & 26.76 & 31.46&     \\
 \nimolo & 26.90 & 31.10 & 31.05  \\
\end{tabular}
\end{ruledtabular}
\end{table}
 To determine qualitatively the nature of bonding between Ni and the
As shell, we have integrated the charge density inside  non-overlapping 
spheres of arbitrary radii around each atom .  These values in units of e 
are given in Table \ref{table3} and show a very small amount of
charge transfer upon formation of the onion-like molecule especially around 
the outer shell As atoms.  The same feature is seen in various charged states 
of the molecule. There could be two possible causes for the small charge
transfer between the atoms. First, the interaction between the \ascage and
\nicage cages may be  weak and  secondly, the environment around  the 
outer As atoms may be  similar in \ascage and in \nimolo.

The strength of interaction between the two cages
will also be reflected in the 
vibrational modes of \nimolo and its infrared (IR) and Raman activity.
If the bonding between the cages is weak, then the  vibrational modes
of the \nimolo will be quite similar to those of isolated \ascage and \nicage.

\begin{figure}[t]
\epsfig{file=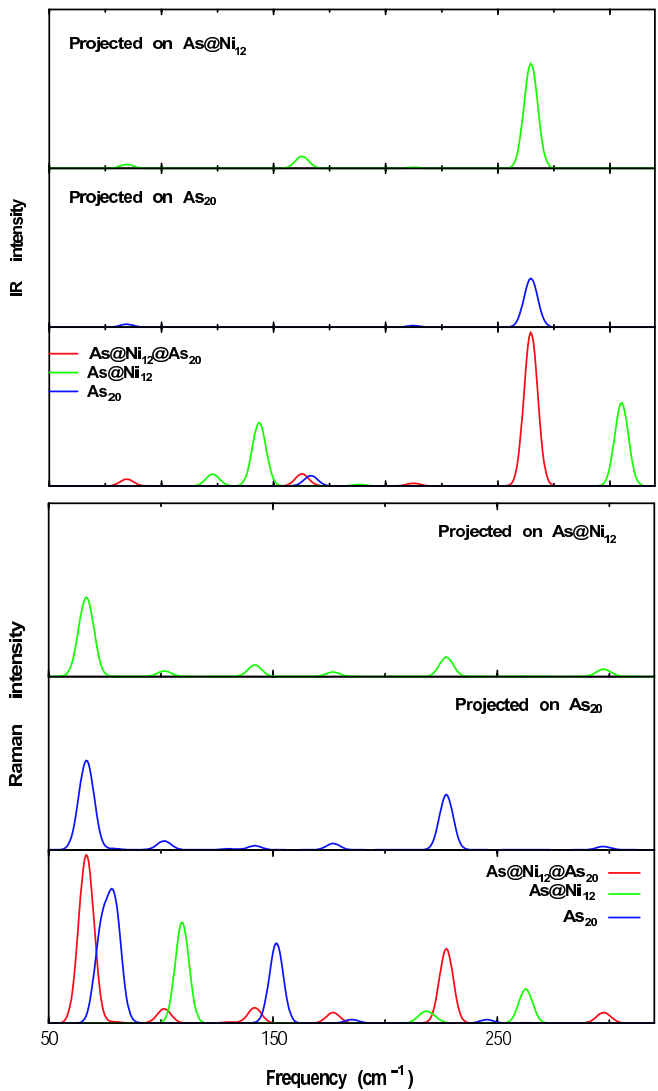,width=\linewidth,clip=true}
\caption{\label{IR} The IR and Raman spectra for the 
\nimolo molecule and its sub-units (see text for details).  }
\end{figure}

The vibrational frequencies for \nimolo and both of its sub-units 
are real indicating that they are  vibrationally stable.
The vibrational frequencies of this massive molecule
range between 42 to 305 \cmin\!.  The
frequencies of the \ascage  and \nicage range from $63 - 248$ \cmin and 
from $104 - 305$ \cmin\!, respectively. 
The mean
polarizability of \nimolo is 25.7 bohr$^3$ per atom which indicates that this
molecule may be useful for application in optical devices. From the
polarizability to molecular volume ratio, we find that the polarizability
of the \nimolo is about 10\% larger than in C$_{60}$ molecule.
 The IR absorption (top three panels) and the 
Raman spectra (bottom three panels) of \nimolo together with its 
isolated sub-units are presented in Fig. \ref{IR}.
The projection of the spectra of \nimolo on the  As and Ni cages are also
shown in the same figure.
The \nimolo has ten optical modes - four IR and six Raman active modes.
These modes and their corresponding symmetries are tabulated in Table 
\ref{table4}.
\begin{table}[b]
\caption{
The IR and Raman active frequencies of \nimolo.
The values are in \cmin\!.
}
 \label{table4}
\begin{ruledtabular}
\begin{tabular}{lllll}
  Symmetry & Frequency \\
\hline
 A$_{1g}$ & 142, 227     \\
 T$_{1u}$ & 84, 163, 212, 265     \\
 H$_{g}$  & 67, 101, 177, 297     \\

\end{tabular}
\end{ruledtabular}
\end{table}

The triply degenerate IR active modes belong to the T$_{1u}$ symmetry.
The IR spectrum of \nimolo is dominated by a large peak
at 265 \cmin with three smaller peaks at 84, 163, and 212 \cmin\!. The
\ascage cage shows little IR activity with a single peak at 167 \cmin
which disappears upon encapsulation.  The low frequency
IR mode of \nimolo at 84 \cmin appears to be due to the rocking motion of the \nicage
in which the \ascage cage also gets mildly distorted. This rocking motion
is also responsible for the peak seen at 143 \cmin in the IR spectra of 
\nicage. The small peak at 163 \cmin in the IR of \nimolo is due to the motion of the 
central As atom. The most dominant peak in the spectrum of \nimolo at
265 \cmin is caused by torsional motion mostly of the Ni cage. This peak
can not be correlated with the peak at 305 \cmin in the \nicage spectra. 

The Raman active modes have A$_{1g}$ and H$_g$ symmetry which are one and five-fold
degenerate. 
The lowest modes of the \ascage and \nicage couple 
and form the two peaks at 67 \cmin and 102 \cmin upon encapsulation.
The increased intensity of the lowest Raman peak 
is a result of increase in polarizability derivative upon encapsulation. In fact,
we find that the ratio of the Raman scattering intensities of \nimolo and 
\ascage scales {\em exactly} as the square of the ratio of their radii.
Similar agreement between the intensities and the ratios of the radii is 
also observed for \nimolo and \nicage. 
These modes
are 5-fold symmetric involving stretching of the molecule diameter. The
mode at 142 \cmin is a breathing mode and correlates to the breathing
mode of \ascage at 152 \cmin. The \nicage shows a Raman peak due to breathing
motion at 263 \cmin. This mode and the breathing mode of the \ascage cage
at 152 \cmin couple together to form a peak at 227 \cmin which is basically
an asynchronous breathing  motion with motions of the Ni and As cage
being in opposite phase. 

The absence of a one-to-one mapping  between the vibrational modes of 
\nimolo and its sub-units in the IR and Raman spectra
indicates that the interaction between these sub-units is non-negligible.
We do not see any  pure ``rattling" motion of \nicage inside the 
\ascage cage.  On the contrary, there seems to be a strong coupling of the 
vibrational modes between the \nicage and the \ascage cages.

\begin{figure}
\epsfig{file=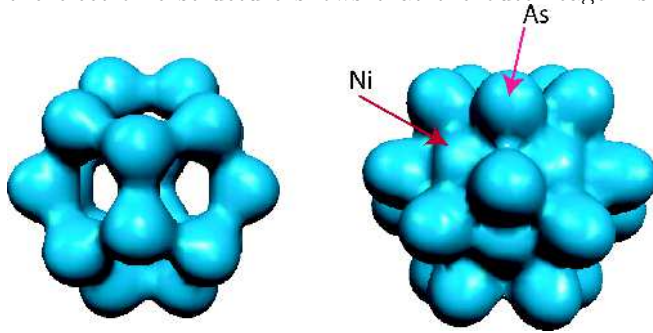,width=\linewidth,clip=true}
\caption{\label{density} The iso-surface of the total charge density 
in \ascage and in \nimolo cluster.}
\end{figure}
The same feature of strong interaction is also seen in the isodensity surfaces
of the \nimolo cluster. In Fig. \ref{density}, we present the isovalued surfaces
of charge density in the isolated \ascage and in the encapsulated \nimolo 
cluster. The density surface in \ascage shows covalent bonding between the 
As atoms. The As atoms in \ascage are sp$^3$ hybridized and form covalent 
$\sigma$-bonds with its three nearest As neighbours. The fourth orbital contains
a lone pair of electrons  pointing radially outward. However, in the 
encapsulated form, each of the outer As atoms have three nearest neighbor
Ni atoms and form strong covalent bonds with the {\it Ni d orbitals rather than
with other As atoms}. We refer to this phenomenon as rebonding of the As atoms.
Thus each As atom has 8 surrounding valence electrons and 
each of the Ni atoms form such covalent bonds with its five nearest As atoms. 
The Ni atoms occupy a position near the center of the pentagonal faces but 
slightly out of plane and since the As-Ni bonds are stronger than Ni-Ni bonds, 
the Ni cage expands.  The expansion of the As cage occurs possibly due to the 
kinetic energy repulsion. It may be mentioned here that the most favorable 
dissociation channel for isolated \ascage is As$_4$ \cite{As20} whereas the 
laser desorption mass spectroscopy shows peaks for all As@Ni12@As$_x$ for x=1,20 
\cite{MFE}.  Since in the encapsulated structure, each As is now bonded strongly 
with three nearest Ni atoms and the As-As bonds are much weaker than in \ascage,
the laser desorption experiment breaks away each As atom individually
rather than in As$_4$.

In conclusion, we find that the isolated \nimolo cluster has I$_h$ symmetry  and is 
vibrationally stable. We find that the molecule will be electronically 
stable in the $-3$ charged state in the  condensed phase due to the opening of a
large HOMO-LUMO gap.  The electrostatic repulsion makes it more stable in 
the $-2$ charged state when in the gas phase.  Our analysis of the IR and Raman 
spectra of the molecule {\it vis-\'a-vis} its sub-units and the electronic 
structure shows that the outer cage As-As bonding changes dramatically upon 
encapsulation of \nicage.  There exists  strong covalent bonding between 
the outer As atoms and inner Ni atoms, in contrast to the Huckel-based 
prediction of Ref. \onlinecite{MFE}.  
Our study provides an understanding of the bonding and will be useful
for further experimental characterization.

  TB and MRP acknowledge financial support from ONR (Grant No. N000140211046) and
by the DoD High Performance Computing CHSSI Program. RRZ thanks GMU
for support and SLR is grateful for financial support as a Distinguished Summer 
Faculty Fellow in the ONR-American Society for Engineering Education Summer Faculty Program.

\end{document}